Das, Anup Kumar (2014). **The 7 Habits of Highly Effective Research Communicators**. In Gautam Maity et. al. (Eds.), *Charaibeti: Golden Jubilee Commemorative Volume* (pp. 356-365). Kolkata, India: Department of Library and Information Science, Jadavpur University. ISBN: 978-81-929886-0-3.

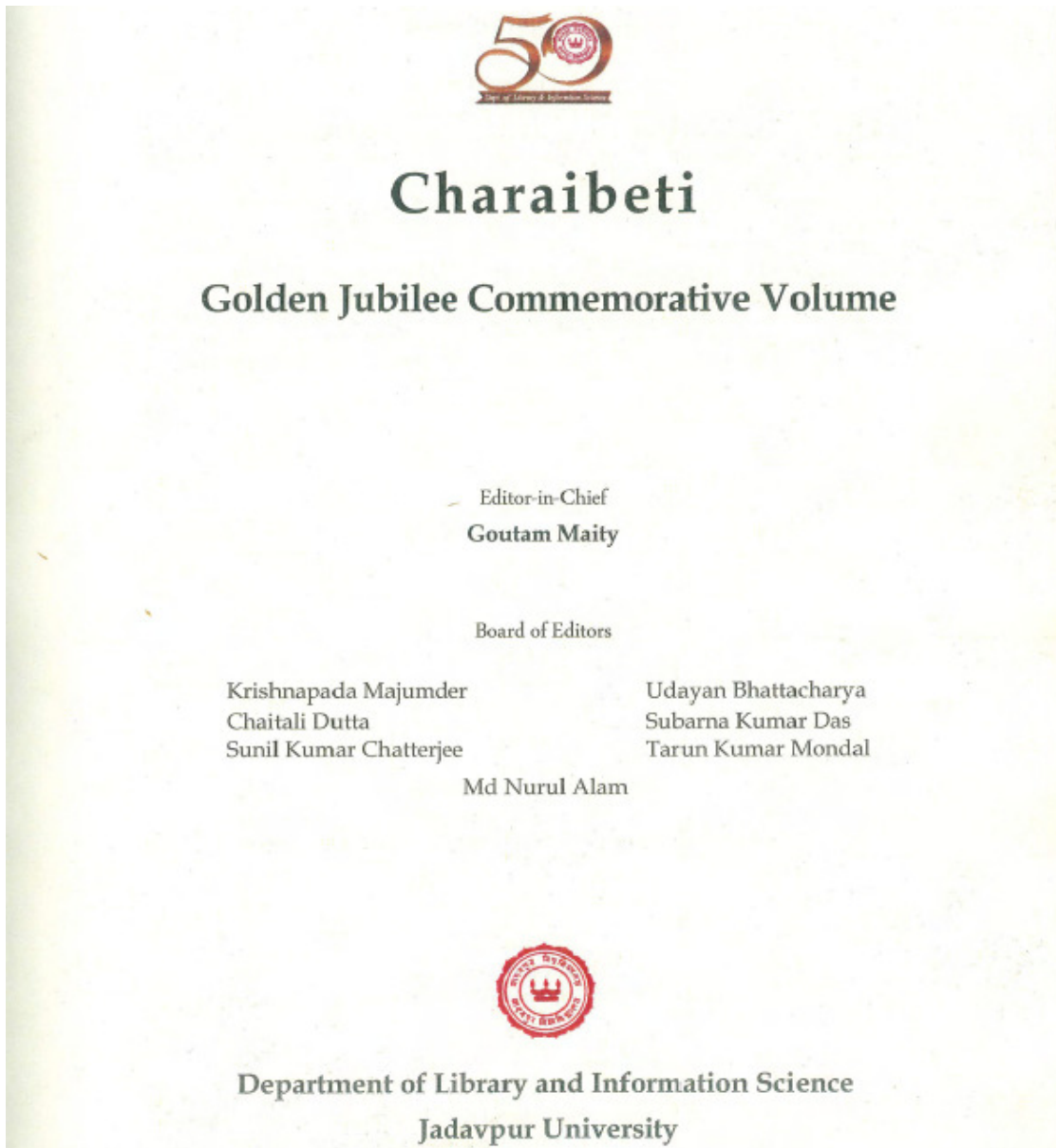

# Charaibeti

## Golden Jubilee Commemorative Volume

Editor-in-Chief
**Goutam Maity**

Board of Editors

Krishnapada Majumder  Udayan Bhattacharya
Chaitali Dutta  Subarna Kumar Das
Sunil Kumar Chatterjee  Tarun Kumar Mondal

Md Nurul Alam

Department of Library and Information Science
Jadavpur University

# The 7 Habits of Highly Effective Research Communicators

**Anup Kumar Das**
Centre for Studies in Science Policy
Jawaharlal Nehru University
New Delhi – 110067, India
http://anupkumardas.blogspot.in

**Abstract**: *The emergence of Web 2.0 and simultaneously Library 2.0 platforms has helped the library and information professionals to outreach to new audiences beyond their physical boundaries. In a globalized society, information becomes very useful resource for socio-economic empowerment of marginalized communities, economic prosperity of common citizens, and knowledge enrichment of liberated minds. Scholarly information becomes both developmental and functional for researchers working towards advancement of knowledge. We must recognize a relay of information flow and information ecology while pursuing scholarly research. Published scholarly literatures we consult that help us in creation of new knowledge. Similarly, our published scholarly works should be outreached to future researchers for regeneration of next dimension of knowledge. Fortunately, present day research communicators have many freely available personalized digital tools to outreach to globalized research audiences having similar research interests. These tools and techniques, already adopted by many researchers in different subject areas across the world, should be enthusiastically utilized by LIS researchers in South Asia for global dissemination of their scholarly research works. This newly found enthusiasm will soon become integral part of the positive habits and cultural practices of research communicators in LIS domain.*

**Keywords**: ResearcherID, Academic Social Network, Social Media, Discussion Forum, Web 2.0, Librarian 2.0, Digital Researcher, Science Communicators

**Type of Article**: Perspective Paper

## I) The 7 Habits

To stay productive and competitive, individual researchers must commit to transforming themselves into full digital communicators. Here are the seven habits that successful research communicators share:

1. ***Create your unique author ID, an identifier for global researchers engaged in academic research***. You may use ResearcherID.com or ORCID.org or both for generating your unique identifier.
2. ***Create your own researcher profile in a dedicated website or personalized webpage and provide up-to-date information about your scholarly research works.*** This will include list of your publications, presentations, research awards, recognitions, travel grants, research projects, latest published works, and professional or creative credentials. You may also create your personalized profile in a blog, using services of Blogger.com or Wordpress.com.
3. ***Create a researcher profile in academic social networks***. Examples of most popular academic social networks are Academia.edu, ResearchGate.net, getCITED.org, SSRN.com, SlideShare.net, Linkedin.com, SkillShare.com, etc. Majority of these academic social networks have provisions of self-archiving of research papers by their registered users. Your deposited works are searchable within these platforms and also from the academic search engines such as Google Scholar, or Microsoft Academic Search, or BASE: Bielefeld Academic Search Engine.

4. ***Share your published works in OA repositories, and also in academic social networks.*** Subject-specific OA repositories and institutional repositories are effective points of dissemination for published scholarly works. In the LIS field, E-LIS Repository (E-LIS: Eprints in Library and Information Science, Eprints.rclis.org) is very effective OA subject repository, similar to arXiv.org in physical and computer sciences. Librarian's Digital Library, maintained by DRTC, is an example of a national level disciplinary repository in the LIS field. There are also online services for cross searching OA knowledge repositories at the global level. One such service is the OAIster (OAIster.org), maintained by OCLC Inc. It indexes metadata from institutional, disciplinary and national-level OA knowledge repositories across the globe. If you have self-archived any of your published work in any OA repository, you can retrieve the relevant data from OAIster database. Similarly, share your delivered lecture notes and lecture slides with Slideshare.net or Speakerdeck.com. Self-archiving is also possible in academic social networks by their registered users. Your deposited works in OA repositories or academic social networks are searchable from the academic search engines.

5. ***Create your profile in Google Scholar Citations and regularly track citations of your published papers***. Google Scholar Citations (GSC) shows your h-index and i10-index scores, list of your publications and number of citations each one received. You need to register for GSC profile, upload your list of publications and make your GSC profile public. Similarly, also create your profile in ImpactStory.org for knowing research impact of your published papers, presentations or shared research data.

6. ***Participate in email-based discussion forums in your specialized area and discuss your research ideas or works in progress***. Works in progress is a set of ongoing research projects. You may also promote published papers by circulating bibliographic information of your recently published papers. Share complete bibliographic information with abstracts and mention its DOI or URI. Don't attach full-text contents with your email, only correct URI or DOI can assist the interested members to locate your papers.

7. ***Make use of free online citation and reference managers.*** You may use EndNote Web, Mendeley, CiteULike, Zotero, Google Scholar Citations Library, or ProQuest's Flow, or combination of some of them, for generating and sharing your reference lists and subject bibliographies. Save and share your references online for accessing anytime and anywhere. Your references, citations and sometimes full-text contents are accessible in cloud platforms through online reference managers.

## II) Brief Description of Related Resources

In this section, you will learn more about different resources mentioned as facilitators of your 7 habits. All of the online resources described here are freely available to researchers in all disciplines. Table 1 indicates comparative features of available authors' unique identifiers, namely ResearcherID and ORCID. Table 2 indicates comparative features of academic social networks commonly used by social science researchers across the world. These social networks complement much popular social networks such as Facebook, Twitter and Linkedin. These are also widely used by LIS professionals and LIS researchers across the globe. However, in this section only some popular academic social networks are briefly discussed to encourage the readers for their early participation and knowledge sharing efforts.

Das, Anup Kumar (2014). The 7 Habits of Highly Effective Research Communicators. In G. Maity et. al. (Eds.), *Charaibeti: Golden Jubilee Commemorative Volume* (pp. 356-365). Kolkata: Department of Library & Information Science, Jadavpur University. ISBN: 9788192988603.   *2*

In the last part of this section, some popular online reference managers and PDF organizers are briefly discussed.

**Table 1: Comparative Features of Authors' Identifiers (IDs)**

|  | **ResearcherID.com** | **ORCID.org** |
|---|---|---|
| **Author's Unique Identifier** | Yes | Yes |
| **Author's Public Profile** | Yes | Yes |
| **Interlinking** | To ORCID & Scopus Author ID | To ResearcherID, Scopus AuthorID & personal website |
| **Sample ID** | A-3366-2012 | 0000-0001-6428-8988 |
| **Profile URL** | www.researcherid.com/rid/A-3366-2012 | http://orcid.org/0000-0001-6428-8988 |

a) **ResearcherID.com:** The ResearcherID.com, an online service offered by Thomson Reuters, is a web-based global registry of authors and researchers. A researcher can freely create a unique identifier called ResearcherID in this system, which is permanent in nature. This ID can be added to publishers' databases during the process of manuscripts submission for uniquely identifying him/her as a contributor. In addition to becoming part of an authors' registry, the researcher can create a public profile and add his/her publication list from the Web of Science (WoS) database or RIS file. While publication list is available in a researcher's public profile, certain citation metrics, citing articles network and collaboration network can also be visible to the profile owner as well as other users searching profiles in this online registry. ResearcherID profile helps in tracking citation count, average citations and h-index of an author from WoS database. Thus, this website becomes very useful tool for author-level measurement if a researcher has good number of papers in his/her credit, which are indexed in WoS database. Figure 1 indicates the basic functions of ResearcherID Registry and how a researcher can obtain citation metrics and analyse impact of research works while making his/her profile public. ResearcherID.com website offers a number of useful features and benefits to researchers as indicated in Text Box 1. ResearcherID profile can also be interlinked to your ORCID profile as well as to your online reference manager account available with the EndNoteBasic (www.myendnoteweb.com), a freely available tool for academic researchers. ResearcherID also helps in exporting bibliographic data to your EndNoteBasic account from ResearcherID profile and stores in your publication list. Here, you can store up to 50,000 bibliographic references with your account, which will help you to prepare bibliographies on different topics or authors for ongoing or future research.

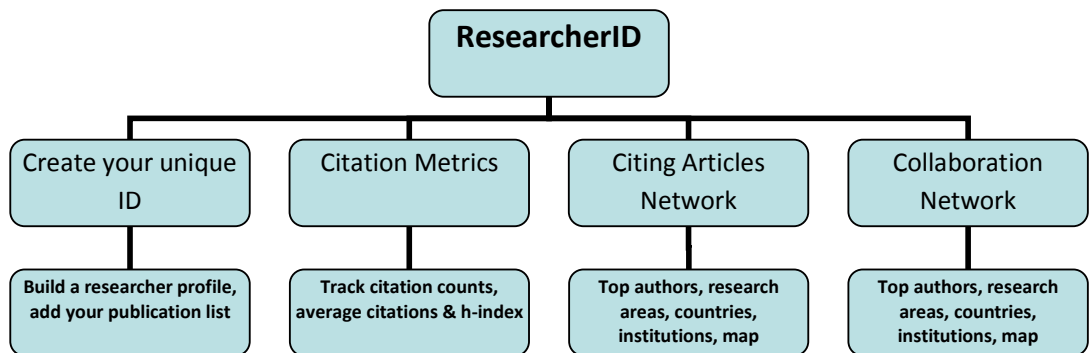

**Figure 1: Basic Functions of ResearcherID Registry**

## Text Box 1: Main Features and Benefits of ResearcherID.com

**Benefits**
- Create a custom profile, choosing what information is public or private.
- Build an online publication list using Web of Science search services, the EndNote Basic online search, or by uploading RIS files.
- Manage your ResearcherID publication list with EndNote Basic.
- Generate citation metrics with times cited information for items added from Web of Science.
- Get links to full text for items added from Web of Science (subject to your subscriptions to full text).
- Add past institution affiliations to your profile.
- Explore the world of research with an interactive map that can help locate researchers by country and topic, or use the new country tag cloud.
- ResearcherID can automatically track times cited counts and citation metrics for records found in Web of Science. Add your publications directly from Web of Science searches.

**Features**
- ResearcherID Badge: Advertise a member's ResearcherID profile on your Web page or Blog. The Badge creates a hovering display of recent publications, and allows viewers to also link to the member's full profile in ResearcherID.
- Collaboration Network: Visually explore who the researcher is collaborating with.
- Citing Articles Network: Visually explore citation relationships based on Web of Science data.

Source: http://thomsonreuters.com/researcherid/

## Text Box 2: ORCID's Mission, Principles and Steps for Creation of an ORCID iD

**Mission**

ORCID aims to solve the name ambiguity problem in research and scholarly communications by creating a central registry of unique identifiers for individual researchers and an open and transparent linking mechanism between ORCID and other current researcher ID schemes. These identifiers, and the relationships among them, can be linked to the researcher's output to enhance the scientific discovery process and to improve the efficiency of research funding and collaboration within the research community.

**Principles**
- ORCID will work to support the creation of a permanent, clear and unambiguous record of research and scholarly communication by enabling reliable attribution of authors and contributors.
- ORCID will transcend discipline, geographic, national and institutional, boundaries.
- Participation in ORCID is open to any organization that has an interest in research and scholarly communications.
- Access to ORCID services will be based on transparent and non-discriminatory terms posted on the ORCID website.
- Researchers will be able to create, edit, and maintain an ORCID identifier and record free of charge.
- Researchers will control the defined privacy settings of their own ORCID record data.
- All data contributed to ORCID by researchers or claimed by them will be available in standard formats for free download (subject to the researchers' own privacy settings) that is updated once a year and released under a CC0 waiver.
- All software developed by ORCID will be publicly released under an Open Source Software license approved by the Open Source Initiative. For the software it adopts, ORCID will prefer Open Source.
- ORCID identifiers and record data (subject to privacy settings) will be made available via a combination of no charge and for a fee APIs and services. Any fees will be set to ensure the sustainability of ORCID as a not-for-profit, charitable organization focused on the long-term persistence of the ORCID system.
- ORCID will be governed by representatives from a broad cross-section of stakeholders, the majority of whom are not-for-profit, and will strive for maximal transparency by publicly posting summaries of all board meetings and annual financial reports.

**Distinguish Yourself in Three Easy Steps**
1. *Register*: Get your unique ORCID identifier Register now! Registration takes 30 seconds.
2. *Add Your Info*: Enhance your ORCID record with your professional information and link to your other identifiers (such as Scopus or ResearcherID or LinkedIn).
3. *Use Your ORCID ID*: Include your ORCID identifier on your Webpage, when you submit publications, apply for grants, and in any research workflow to ensure you get credit for your work.

Source: http://orcid.org/about/

Das, Anup Kumar (2014). *The 7 Habits of Highly Effective Research Communicators*. In G. Maity et. al. (Eds.), *Charaibeti: Golden Jubilee Commemorative Volume* (pp. 356-365). Kolkata: Department of Library & Information Science, Jadavpur University. ISBN: 9788192988603.    *4*

## Table 2: Major Academic Social Networks

|  | **Academia.edu** | **ResearchGate.net** | **getCITED.org** | **SSRN.com** |
|---|---|---|---|---|
| **Target Group** | Academics: researchers, students | Researchers | Researchers | Researchers, Authors |
| **Subject Coverage** | All | All | All | Social Sciences, Humanities and Law |
| **Founded in** | 2008 | 2008 | 2004 | 1994 |
| **Mission** | To accelerate the world's research; to make science faster and more open. | To give science back to the people who make it happen and to help researchers build reputation and accelerate scientific progress. | To make records of scholarly work publicly available. | To provide rapid worldwide distribution of research to authors and their readers and to facilitate communication among them at the lowest possible cost. |
| **Web 2.0 Interactivity** | Yes | Yes | No | No |

b) **ORCID.org:** ORCID stands for Open Researcher and Contributor ID. Similar to ResearcherID.com, ORCID.org provides a persistent digital identifier that distinguishes a researcher from every other researcher. Creation of an ORCID iD for a researcher is very easy and free. Here, you have to provide certain personal and professional details to include your name in a registry of unique researcher identifiers. After successful registration, a unique ORCID iD is generated and a user profile is created in the website. Example of an ID is 0000-0001-6428-8988, which includes 16 numerical characters and three separators. You can integrate your other profiles or unique author's identifiers available elsewhere such as ResearcherID, Scopus and LinkedIn. Your publication list will also be added to your profile, which includes bibliographic information of published scholarly works and hyperlink to full-text contents of each work. A publication list can be obtained from Scopus database that will include bibliographic record of papers published in Scopus-covered journals. Other relevant works can also be added in your profile through importing bibliographic data from a RIS file of your list of publications. ORCID.org maintains a searchable registry of researchers that helps in identifying researchers from your field, from an institution, collaborator, city or country. Funding agencies also can keep track on researchers' works, funded by them or considering funding in near future. The website provides APIs that support system-to-system communication and authentication to online systems of funders, publishers and others that require ORCID identifiers.

c) **Academia.edu:** This is one of the largest social networking websites for academics. Established in 2008, it is a social media space for academics and researchers to make their academic works visible to global communities of academicians and researchers. Any student, a researcher or a faculty member from any subject area, affiliated to a higher educational institution or a university, can freely create a profile and upload his/her published or unpublished papers, conference presentations and research datasets for worldwide dissemination. The researcher here has options to upload full-text contents, or to provide only bibliographic details. He/she can seek academic collaborations, professional advice and feedbacks from fellow network members. One may follow a number of researchers and peers. Many of the persons one *Following* are either his/her mentors, fellow researchers, colleagues, peers, supervisors, teachers, collaborators and co-authors. Higher number of Followers of a profile indicates that researcher's research works get considerable attention to researchers in his/her domain and adding value to the volumes of current research literature. Academia.edu offers four basic functionalities: (i) Share your papers, (ii) See analytics on your profile and papers, (iii) Follow other people in your field, and (iv) Interact with other researchers. A public profile in this website provides some indicators on popularity of the profile holder, namely, number of profile views, number of document views, number of self-archived papers, number of followers, number of followings and number of forum posts.

d) **ResearchGate.Net:** This is one of the most prominent professional networks for scientists and researchers. Established in 2008, it is a social media space for researchers to make their research visible to global researchers' communities. Any researcher from any subject area can freely create researcher's profile and upload their published, unpublished, working papers and research datasets for worldwide dissemination. The researcher here has options to upload full-text contents, or to provide only bibliographic details. He/she can also add details of his/her completed and ongoing research projects for further discussions, dialogues and collaborations with network members. As a registered member in this online platform, you can read the latest publications in your field shared by other fellow researchers; discuss your work with other specialists; and collaborate with colleagues located in the same country or other countries around the world. A researcher's profile provides statistics related to his/her research works, such as number of papers available, total publication views, total full-text downloads, total dataset downloads, total full-text requests, citations. Your profile also indicates number of *Followers* you have, number of researchers you are *Following*, and *Top Co-authors*. ResearchGate generates *RG Score* for every registered researcher. The RG Score is a metric that measures scientific reputation based on how all of your research is received by your peers. It is a mix of indicators based on statistics related to your publications, questions, answers and followers. RG Score is a derived in combination of publications (their views, downloads & citations), questions & answers (interactions with other members) and number of followers. For example, a member's RG Score is shown as 40.73, while he authored 358 publications with 777 citations and 529.24 impact points.

e) **getCITED.org:** This is a registration-based website for facilitating academic communities in sharing bibliographic information on published and unpublished academic papers and other documents. Established in 2005, it has become a social space for academics and researchers to make their academic works visible to global communities of academicians and researchers. Any researcher or a faculty member from any subject area, affiliated to a higher educational institution, research institution or university, can freely create a profile and upload his/her list of publications. A registered user can add bibliographic details of his/her published or unpublished papers, books, book chapters, theses, dissertations, conference presentations, reports and other documents for increasing their worldwide visibility. However, this site does not have facility of uploading full-text contents. A researcher's profile in getCITED.org includes profile statistics, such as, number of publications in each category, citation rank, researcher's rank, number of citations, number of views of his/ her this profile. This website performs more as a repository of bibliographic contents than an academic social network. This platform does not facilitate social networking with other members of and sharing knowledge products within the academic communities.

f) **SSRN.com:** The Social Science Research Network (SSRN.com) is a document repository for worldwide dissemination of social science research. It comprises of about 22 specialized research networks in many of the specialized domains of social sciences, humanities and law. Individuals, institutions, publishers and scientific societies can share their publications and other academic contents for global dissemination through this single gateway. This website was launched in 1993 and is presently owned by the Social Science Electronic Publishing Inc., based in the United States. Its individual and institutional members spread around the world have made this website one of the top-ranking digital repositories with significant amount of open access contents. The SSRN website secured second position in the 14$^{th}$ edition of the World Ranking Web of Repositories (http://repositories.webometrics.info/en/world), which was announced in January 2014. SSRN has a unique "Partners in Publishing" program and it works with over 1,800 scientific journals and research institutions. These partners provide information on forthcoming papers and permission to have their work posted to SSRN. SSRN aggregates working papers from many leading institutions and think tanks. Each registered individual member is free to upload his/her published papers and other academic contents and disseminate to global researchers communities. Full-text contents submitted by an author for global dissemination can be of either open access or out of any copyright-restriction. However, an author's briefcase or workspace usually displays papers in four categories: (i) Publicly available papers, (ii) In process papers, (iii) Privately available papers, and (iv) Inactive papers. Only papers in category (i) are available in the SSRN eLibrary. eLibrary papers are searchable from SSRN portal and by external academic search engines.

g) **DOI and URI:** Digital Object Identifier (DOI) is a persistent resource locator for an online scholarly content published in academic journals, conference proceedings and e-books as an article, or a conference paper, or a book chapter. There is always a possibility of getting a dead link or broken link to a published scholarly article, while publishers have moved their website or webpages due to website maintenance or up-gradation. In order to facilitate retrieval of each scholarly item, a DOI is assigned to every online item. This reduces possibility of getting a broken link in future. For example, an article has assigned its DOI as *10.4018/978-1-4666-4365-9.ch002*. This article can be retrieved using following web address: *http://dx.doi.org/10.4018/978-1-4666-4365-9.ch002*. Then this address will redirect browser to a publisher's article page, say *http://www.igi-global.com/chapter/introduction/103067*, where that full-text scholarly content is presently available. Uniform Resource Identifier (URI) is a string of characters used to identify a name of a web resource. In above-mentioned example, *http://www.igi-global.com/chapter/introduction/103067* is the absolute URI.

h) **Google Scholar Citations:** This web service, offered by the Google Corporation since 2012 from the site *http://scholar.google.co.in/citations*, helps individual academic researchers in highlighting their performance indicators based on citations received by their published scholarly works. Here individual scholars, logging on through a Google account, can now create their own page giving their fields of interest and citations. Google Scholar automatically calculates and displays the individual's total citation count, h-index and i10-index. Researcher's public profile also displays institutional affiliation, research interests, hyperlink to researcher's homepage, and citations history. An example of a researcher's public profile (of Professor Amartya Sen) is *http://scholar.google.com/citations?user=sLNFo0sAAAAJ*.

i) **E-LIS Repository:** The Eprints in Library and Information Science (E-LIS) (http://eprints.rclis.org) is an international open access repository for academic papers in library and information science (LIS). Its website facilitates self-archiving by the registered users, who are usually combination of young and experienced LIS researchers working across the globe. Launched in 2003, this open access disciplinary repository has become most resourceful repository for retrieving LIS literature with multilingual contents, covering literatures written in 22+ world languages.

j)  **ImpactStory.org:** This is a leading provider of article level metrics data, helping individual researchers to know impacts of their published works. This website offers registered users creating their impact profile on the web, revealing diverse impacts of their articles, datasets, software, and presentations. This is a collaborative not-for-profit open source project supported by the U.S. National Science Foundation (NSF), Alfred P. Sloan Foundation and Open Society Foundation. ImpactStory.org helps in creating author's profile and adding publication list through importing bibliographic records from different sources such as Scopus database, ORCID.org, Google Scholar Citations, SlideShare and many others. A researcher can create a profile for free in this website to know how many times his/her work has been downloaded, bookmarked, and blogged. A researcher can also generate code to embed ImpactStory profile into his/her institutional CV and research blog.

k)  **Discussion Forums and Mailing Lists:** Online discussion forums play significant roles in an early career researcher's life. Some of the online forums are email-based mailing lists, while some others are e-groups in social networking platforms. They provide platform of interactions in sharing ideas or seeking expert comments or feedbacks on your research questions, methodologies or research publications. These are also rich sources of information with calls for papers in journals or conferences, and calls for participations in doctoral symposiums or training workshops. The most important and popular ones in India are:
- *LIS Forum* (lis-forum@nsci.iisc.ernet.in), initiated by the NCSI of Indian Institute of Science Bangalore;
- *SALIS e-Group* (salis_info@yahoogroups.com), initiated by the Society for the Advancement of Library and Information Science, SALIS, Chennai;
- *DLRG: Digital Library Research Group* (dlrg@drtc.isibang.ac.in);
- *Digital Libraries: India* (digilib_india@yahoogroups.com);
- *KM-Forum* (km-forum@yahoogroups.com);
- *LIS Leadership Forum* (lislforum@gmail.com),
- *E-Librarian* (elibrarian@googlegroups.com).

In a recent paper "An Exploratory Analysis of Messages on a Prominent LIS Electronic Discussion List from India" authors analysed informative, collaborative and networking roles played by fellow participants in LIS Forum of IISc (Pujar, Mahesh, & Jayakanth, 2014). Another online forum *LIS Links* or, Library and Information Science Links (Lislinks.com) has become India's first and one of the largest social networks for library professionals, which is a platform for about 15000 registered participants.

l)  **Online Citation/ Reference Managers**
- **Mendeley**: It is one of the most preferred online reference managers freely available to researchers across the world. Launched in 2007, it later acquired by Elsevier B.V. which is the owner company of online products Scopus and ScienceDirect. It became leading online reference manager and PDF organizer, in terms of its popularity amongst researchers and academics. Any researcher can create a free online account in Mendeley platform, store bibliographic records as well as full-text documents in PDF or other formats and later retrieve those saved documents as and when required. A personal library of a user can store all downloaded or collected literatures one uses in ongoing, past or forthcoming research projects. Mendeley offers 2GB (gigabytes) personal storage space, where you can store full-text documents, and 100MB (megabytes) shared storage space, where you can share full-text documents of your choice. However, beyond this limit you have options to upgrade to paid value-added services, namely, Mendeley Premium Packages and Mendeley Institutional Edition. It also offers a desktop version, called Mendeley desktop, which is one of the most downloaded free reference manager software. Here a user can organize his/her PDF collections and add annotated notes in each document file.

- **EndNote Basic**: The EndNote is well-known proprietary reference management software widely used by researchers across the world. In December 2006 the brand owner and developer Thomson Reuters launched a web-based version of EndNote, called EndNote Web. In April 2013 Thomson Reuters launched a free version of EndNote Web, called EndNote Basic (Endnote.com/basic/) – available to individual researchers across the world without any subscription obligation. However, EndNote Basic has some limited functionalities as compared to its full-version. Similar to Mendeley and Zotero, EndNote Basic has less storage space to store full-text contents of references in a user's library collection. EndNote Basic can gather bibliographic information using EndNote Bookmarklet from an article page of electronic journals and store this bibliographic record in your personal library available with your user account. EndNote Basic offers you a *Cite While You Write™* plug-in for MS-Word. You can use the EndNote plug-in to insert references, and format citations and bibliographies automatically while you write your papers in Word. This plug-in also allows you to save online references to your personal library in Internet Explorer for Windows environment. Your registered profile in EndNote Web can also be interlinked to your ResearcherID profile for exchanging citations of contributed papers.

- **CiteULike**: CiteULike.org is another most preferred online reference manager freely available to researchers across the world. Launched in 2004, CiteULike became a pioneer in offering services of online reference manager and PDF organizer. It has become very popular since its inception amongst researchers and academics. Any researcher can create a free online account in CiteULike platform, store bibliographic records as well as full-text documents in PDF or other formats and later retrieve those saved documents as per their research requirements.

- **Zotero**: Zotero.org is another online reference manager freely available to researchers across the world. It was launched in 2006 as a project of the Roy Rosenzweig Center for History and New Media, at George Mason University in the United States, with its web-based solution as well as desktop open source application for reference management and PDF organizer. It has become popular amongst researchers and academics in many countries. Any researcher can create a free online account in Zotero platform, store bibliographic records as well as full-text documents in PDF or other formats and later retrieve those saved documents as per their research requirements. Zotero web platform helps a registered user to collect, organize, cite, and synchronize references of scholarly works and collaborate with research groups and online forums for knowledge enrichment. Any researcher can freely download Zotero desktop application for reference management and organizing full-text documents in a desktop environment. Zotero bookmarklet is also available to Zotero users to import a citation from a publisher's article page and save it to a user account in Zotero web.

- **Flow**: Flow.proquest.com is another online reference manager freely available to researchers across the world. Launched in 2013 by the ProQuest Inc. with features similar to existing online reference manager RefWorks® – another from the same company. While RefWorks is a set of priced online services for institutional subscribers, Flow is a free version for individual researchers. It is a powerful collaboration and document management tool. It manages researcher workflows while integrating document management and sharing with citation data. This tool enables registered users to discover and manage content, store and organize documents, and through integration with Microsoft® Word, write papers, supported with instant bibliographies and annotation. selected or chosen

- **Google Scholar Library**: If you are a registered user of Google Scholar Citations or any other Google product, then you can maintain a collection of publications, based on your saved items – called *My library*, your contributed papers – called *My Citations*, and your cited items – called *Cited by me*. Often, you use Google Scholar search engine and retrieve some important references on a given search term. At the bottom of bibliographic description of each retrieved publication, you will find a *Save* option to include that particular item in the *My library* collection maintained by you. The saved items can be retrieved later as and when required. You can also create new labels to classify items in your collection for easy retrieval in the future.

### III) Conclusion

Very often, research communicators, located in the developing countries, are naïve in handling academic communication channels available to them with Web 2.0 readiness. They need to unlearn many traditional scholarly communication processes (web 1.0 or producing non-digital objects) and learn Web 2.0-enabled interactive scholarly communications, and relearn the similar processes with new insights in every technological up gradation. The new-age researchers need to understand and grasp changing landscape of research communications. The successful ones should be open to experimentation. Although the age of experimentation with digital might be over in the enterprise spaces (Olanrewaju, Smaje & Willmott, 2014), reaching out to your intended audiences is very vital as research communicators in developing countries.

### IV) References and Further Readings

5. Pujar, S. M., Mahesh, G., & Jayakanth, F. (2014). An Exploratory Analysis of Messages on a Prominent LIS Electronic Discussion List from India. *DESIDOC Journal of Library & Information Technology*, *34*(1).

**V) List of Abbreviations**

| | |
|---|---|
| API | Application Programming Interface |
| DOI | Digital Object Identifier |
| GB | Gigabytes |
| LIS | Library and Information Science |
| MB | Megabytes |
| ORCID | Open Researcher and Contributor ID |
| PDF | Portable Document Format |
| RIS File | A standardized tag format developed by Research Information Systems (RIS) |
| SSRN | Social Science Research Network |
| URI | Uniform Resource Identifier |
| URL | Uniform Resource Locator |

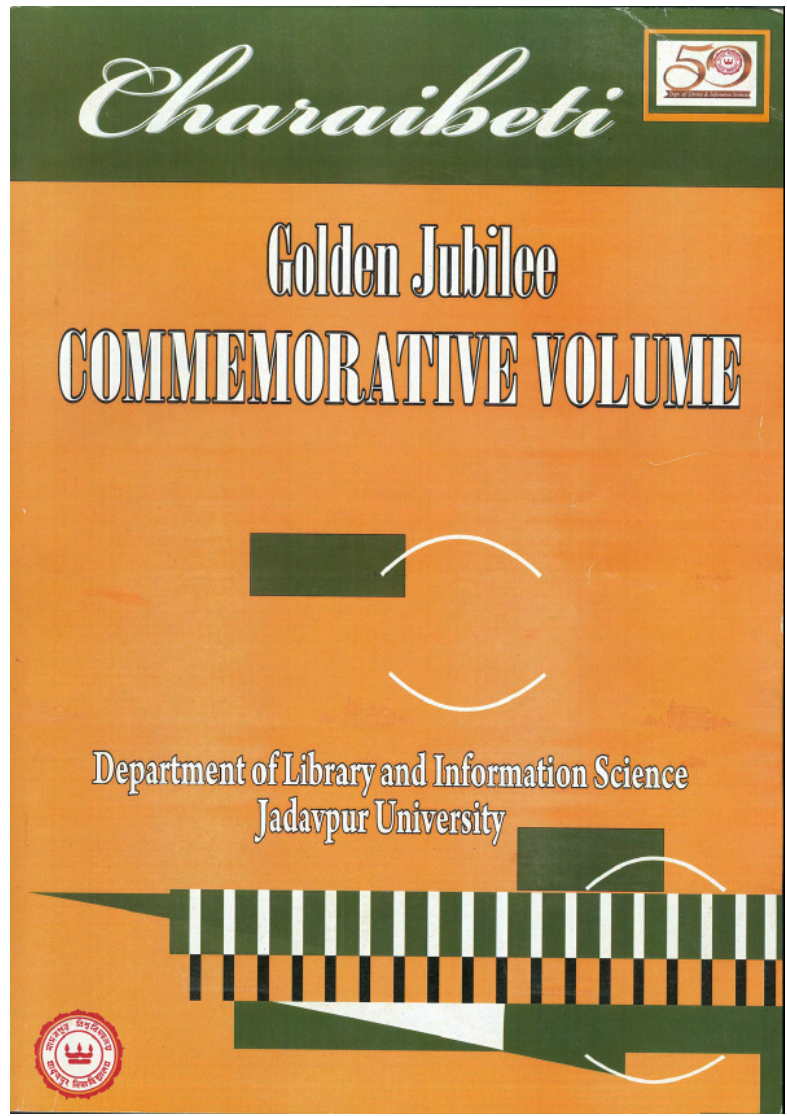